\documentclass[doublecol,comment]{epl2} 

\newcommand{\Jeff}{J^\mathrm{eff}}
\newcommand{\heff}{H^\mathrm{eff}}

\title{Comment on ``Creating artificial magnetic fields for
cold atoms by photon-assisted tunneling'' by Kolovsky A.R.}
\shorttitle{Title} 

\author{C.E. Creffield \and F. Sols }

\institute{                    
Departamento de F\'isica de Materiales, Universidad
Complutense de Madrid, E-28040, Madrid, Spain \\
}

\pacs{67.85.Hj}{Bose-Einstein condensates in optical potentials}
\pacs{03.65.Vf}{Phases: geometric; dynamic or topological}

\begin{document}

\maketitle

Cold atom systems held in optical lattice potentials are proving
to be a remarkably versatile means of investigating coherent
quantum phenomena. A particularly interesting
recent development has been the use of these systems to 
study synthetic gauge fields. In Ref.\cite{kolovsky}
a scheme was proposed to simulate a uniform magnetic field by 
applying a periodic driving to the optical lattice, and employing
the effect known as ``photon assisted tunneling'' to generate
the required Peierls phases. Here, however, we show that the
analysis of this system was incorrect;
contrary to the central claim of the paper,
a driving of the kind described cannot produce a uniform
synthetic magnetic field.

For simplicity, we begin by considering a particle hopping on a
one-dimensional lattice, described by a tight-binding model
\begin{equation}
H_0 = -\frac{J}{2} \sum_{m} \left(| m \rangle \langle m+1 | + H.c.\right) \; ,
\label{1d_hubbard}
\end{equation}
where $| m \rangle$ are Wannier states localized on sites $m$,
and $J$ is the hopping between nearest neighbors. Oscillating
the position of the optical lattice,
$x \rightarrow x + x_0 \cos( \omega t + \phi)$,
produces an inertial force in the rest frame of the lattice
$F_I = - M x_0 \omega^2 \cos( \omega t + \phi)$, which can be
described in terms of a potential function
\begin{equation}
F_I = - \frac{ \partial}{\partial x} V(x,t) \; .
\end{equation}
This allows the driving to be included in the Hamiltonian as
\begin{eqnarray}
H &=& H_0 + V(x,t) \nonumber \\
&=& H_0 + a F_\omega \cos(\omega t + \phi) \sum_m |m \rangle m \langle m | \;,
\end{eqnarray}
where $a$ is the lattice spacing and $F_\omega = M x_0 \omega^2$.
In addition to the oscillation the lattice can also be
subjected to a constant acceleration \cite{morsch_accel}, which adds a
constant component $F$ to the total lattice potential
\begin{eqnarray}
\label{driving}
H &=& -\frac{J}{2} \sum_{m} \left(|m\rangle \langle m+1 | + H.c.\right) \nonumber \\
&+& a \left( F + F_\omega \cos(\omega t + \phi) \right)
\sum_m |m \rangle m \langle m | \;.
\end{eqnarray}
It can be readily seen that this Hamiltonian mimics that of a charged particle 
subjected to a uniform electric field, with 
$F$ and $F_\omega$ playing the roles of the DC and AC components respectively.

When the static component matches the driving frequency, $F = \omega$, 
a straightforward Floquet analysis \cite{cec,kol_pra} reveals
that in the high-frequency limit the dynamics of the driven system can be
described by an effective static Hamiltonian
\begin{eqnarray}
\label{phases}
\heff &=& -\frac{{\Jeff}}{2} 
\sum_m \left(|m+1\rangle \langle m | e^{i\phi} + H.c.\right) \\
\Jeff &=& J{\cal J}_1(F_\omega/\omega) \nonumber \;.
\end{eqnarray}
where ${\cal J}_1(x)$ is the Bessel function of the first kind.
The presence of the phase factors, $e^{\pm i\phi}$, in the effective
Hamiltonian provides the motivation for attempting to use this kind
of driving to generate synthetic gauge potentials.

\bigskip

A two-dimensional square lattice threaded
by a uniform magnetic field can be described in the Landau 
gauge ${\bf A}=A(0,x,0)$ by
\begin{eqnarray}
\label{landau}
H = &-&\frac{J_x}{2} \sum_l \left(|l+1,m\rangle \langle l,m | + H.c.\right) + \nonumber \\
&-&\frac{J_y}{2} \sum_m \left(|l,m+1\rangle \langle l,m | e^{i \phi l} + H.c.\right)
\end{eqnarray}
where  $J_x / J_y$ are the tunneling amplitudes in the $x / y$
directions, and $(l,m)$ are the coordinates of the lattice sites.
In Ref.\cite{kolovsky} it was proposed that this pattern of Peierls phases
could be mimicked by adding the periodic driving potential
\begin{equation}
\label{drive}
V(x,y,t) = y \left( F +  F_\omega \cos( \omega t - x \phi ) \right)
\end{equation}
to the two-dimensional tight-binding model. 
With this driving potential, the force experienced in
the rest frame of the lattice (${\bf F} = - \nabla V $) is given 
in the continuum limit by
\begin{eqnarray}
F_x &=& - \partial_x V  = - \left[
\phi F_\omega \sin \left( \omega t - x \phi \right) \right] y
\label{Fx} \\
F_y &=& - \partial_y V  = - \left[
F + F_\omega \cos \left( \omega t - x \phi \right) \right] \;.
\label{Fy}
\end{eqnarray}
Comparing (\ref{Fy}) with Eqs.\ref{driving} and \ref{phases} indicates
that the tunneling in the $y$-direction will indeed acquire the correct
phases to reproduce the Landau gauge. However, it is also apparent that
tunneling in the $x$-direction will experience a periodic driving
(\ref{Fx}) too, and so will {\em also} be renormalized. 
This may appear unexpected, as this renormalization is occurring
for tunneling perpendicular to the shaking of the lattice, 
but is a consequence of the explicit $x$-dependence of $V(x,y,t)$.
This effect was not considered in
Ref.\cite{kolovsky}, as $F_x$ was implicitly assumed to be equal to zero.
Consequently the results obtained in Ref.\cite{kolovsky} are only valid
for $F_x \simeq 0$,
that is, in a sufficiently small region around $y=0$.

\begin{figure}
\includegraphics[width=0.24\textwidth,clip=true]{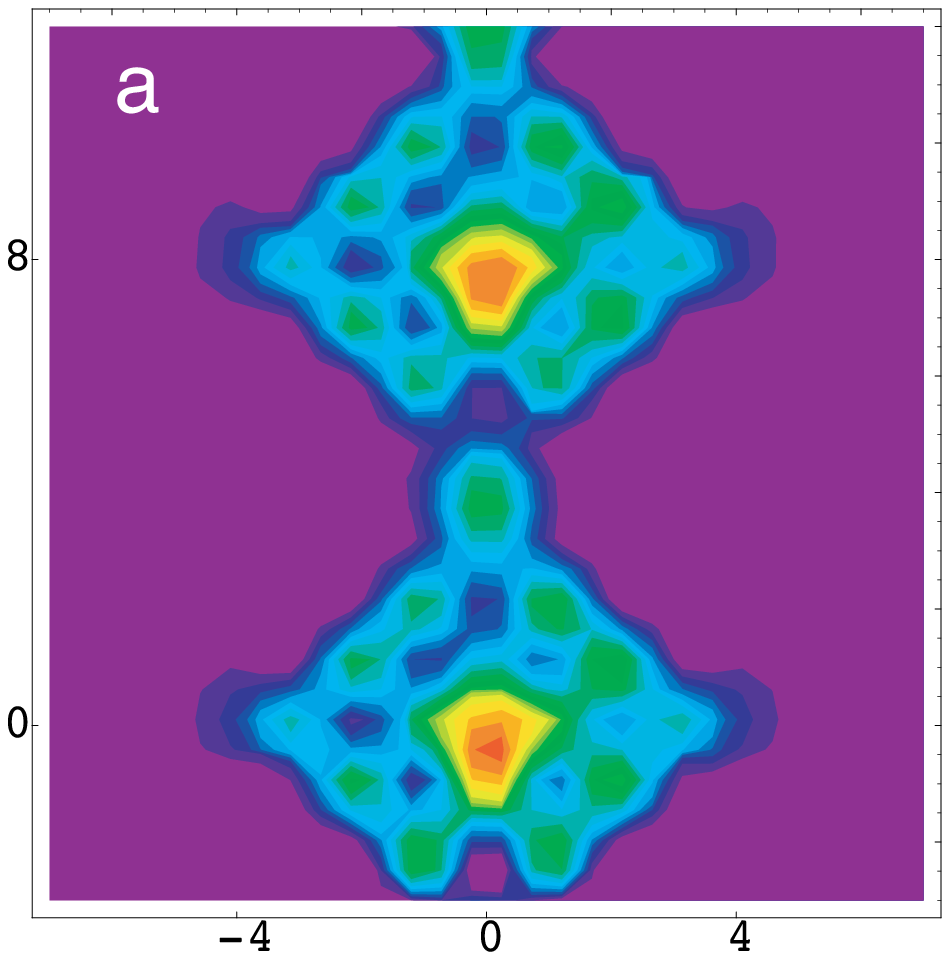}
\includegraphics[width=0.24\textwidth,clip=true]{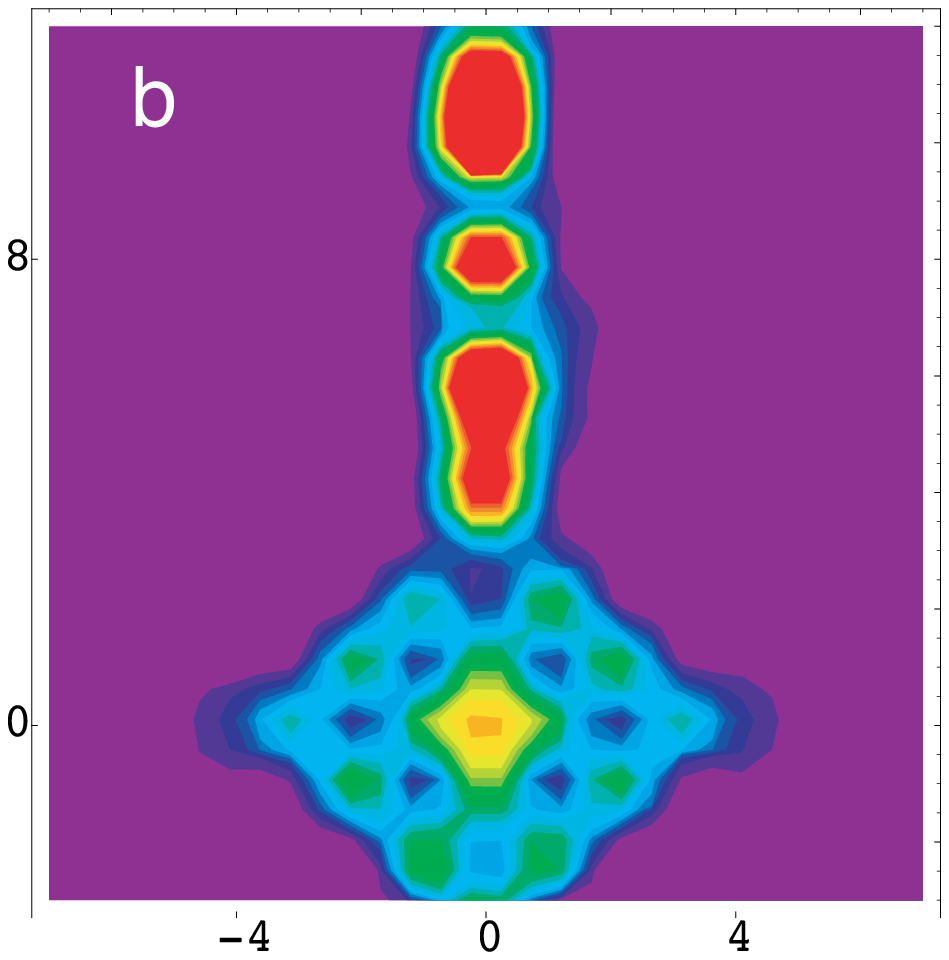}
\caption{Evolution of the initial state $\psi(x,y) =
N \left( \exp[-2 (x^2 + y^2)] + \exp[-2 (x^2 - (y-8)^2)] \right)$.
(a) Density distribution of the system at $t=30T$ after
evolving in the presence of a uniform magnetic
field (Eq.\ref{landau}).
(b) As in (a), but for a periodically-driven system.
The peak centered on
the origin shows a similar evolution to the exact case, but the other
peak is instead compressed into a narrow strip parallel
to the $y$-axis, due to the
renormalization of $J_x$. Parameters are $\omega = 19.9$,
$F/\omega=1$, $F_\omega/\omega = 0.2$, and
$\phi = \pi/3$.}
\label{compare}
\end{figure}

Performing a Floquet analysis reveals that 
while $J_y$ is renormalized as in Eq.\ref{phases}, the 
the effective $x$-hopping is given by
\begin{equation}
\Jeff_x = J_x {\cal J}_0 \left( 2 y \sin \left[\phi/2\right] \  
\frac{F_\omega}{\omega} \right) \; . 
\label{transverse}
\end{equation}
Thus away from $y = 0$,
$J_x$ will be suppressed by the zeroth Bessel function. In Fig.\ref{compare}
we show the magnitude of this effect. The system is initialized in a 
superposition of two narrow Gaussians, one centered on the origin, and the other
on $(0,8)$. Fig.\ref{compare}a shows the system 
evolving under the action of Eq.\ref{landau} for a magnetic
flux of $\phi=\pi/3$, while in Fig.\ref{compare}b we show the result of
evolving the system under the time-dependent driving potential (\ref{drive}). 
We can see that the peak centered on the origin behaves similarly in both cases,
indicating that near this point the driving indeed mimics a magnetic field
with reasonable accuracy.
The behavior of the other peak, however, is strikingly different.
The suppression of tunneling produced by the Bessel function squeezes
the wavepacket into a narrow band, with a very different
dynamics to that of the true magnetic flux.

\bigskip

In summary we have shown that the scheme proposed in Ref.\cite{kolovsky}
cannot produce a uniform magnetic field. At best it can produce a field
that is approximately constant over a finite number of lattice spacings, which
severely limits its applications. We have demonstrated this by an 
explicit calculation of the effect of the driving potential. 
This result is also consistent with the basic observation
that a vector field with a non-zero curl
cannot be described by a conservative potential.
More complicated experimental setups are required to avoid this drawback;
for example introducing additional optical lattice potentials 
\cite{bloch} to compensate for the transverse renormalization (\ref{transverse}).

\acknowledgments
The authors would like to acknowledge the assistance of Andrey Kolovsky
in the preparation of this work. This research was supported by
the Spanish MINECO through Grant No. FIS-2010-21372.

\end{document}